\documentclass{article}

\usepackage{iclr2027_conference,times}
\usepackage[varqu,varl]{zi4}
\usepackage{graphicx}
\usepackage{adjustbox}
\usepackage{booktabs}
\usepackage{amsmath}
\usepackage{amssymb}
\usepackage{enumitem}
\usepackage{xcolor}
\usepackage[hyphens]{url}
\usepackage[hidelinks]{hyperref}
\ExplSyntaxOn\NewDocumentCommand{\task}{m}{\tl_set:Nn\l_tmpa_tl{#1}\tl_replace_all:Nnn\l_tmpa_tl{-}{-\allowbreak}\texttt{\l_tmpa_tl}}\ExplSyntaxOff

\usepackage{xurl}
\iclrfinalcopy
\begin{document}

\title{FDE-Bench: Evaluating LLM Agents for \\ Deployment Environment Configuration}
\author{Weihang Ding$^{1}$, Junfei Zhan$^{2}$, Yueting Li$^{1}$, Qirong Guo$^{3}$ \\ $^{1}$University of California, Berkeley \quad $^{2}$Imperial College London \\ $^{3}$The Hong Kong University of Science and Technology (Guangzhou)}
\maketitle
\lhead{Preprint}

\begin{abstract}
Deployment requires an agent to turn application code into a running system whose services connect, become ready, and remain observable. FDE-Bench evaluates this capability with 136 deployment-configuration tasks spanning Docker images, multi-service Compose stacks, and Kubernetes, in greenfield and diagnose-and-repair modes. Agents submit declarative artifacts that are collected, rebuilt, and redeployed in a pristine environment. Four gated binary check layers measure build, readiness, behavior, and conformance to the deployment specification, using programmatic checks without an LLM judge. A four-arm release gate requires a resolving reference solution and rejects tasks solved by do-nothing, specification-transcription, or generic-stub submissions. The released check annotations expose the link between 2{,}145 checks and their specifications, including seven documented gaps. Three additional adversarial strategies test shortcuts in the grading signals; none resolves any of the 135 tasks they cover, while a vacuous health probe passes readiness and exposes the need for downstream checks. On the 136-task evaluation grid, seven language models from four providers use the same four-tool scaffold and resolve 52.9--75.0 percent of tasks. The three zero-intelligence floors resolve none and reach a mean Deployment Score of at most 0.44. Readiness is the largest failure stage, accounting for 110 of 313 unresolved episodes. Mean resolution rate is 30.7 percentage points higher on the repair task group than on the disjoint greenfield group, with a positive gap for every model; ten tasks resist all seven. In a 25-task case study, one practicing engineer directing Claude-Sonnet-5 resolves 92 percent against 72 percent for the autonomous baseline. FDE-Bench links deployment success and failure to artifacts that can be inspected and replayed.
\end{abstract}

\section{Introduction}
\label{sec:intro}

Repository-level benchmarks ask agents to modify code so that it passes tests in a working environment \citep{jimenez2024swebench,chan2024mlebench}. Deployment additionally requires managing hardware budgets, dependencies, container images, orchestration, service connectivity, and health monitoring. Passing repository tests alone does not establish deployment competence. FDE-Bench\footnote{\url{https://anonymous.4open.science/r/fde-bench-dev-B471/}}, for forward-deployed engineering, evaluates agents tasked with producing a running, healthy, and monitored system.

SWE-bench resolution rose from single digits \citep{jimenez2024swebench} to a majority in roughly two years \citep{swebench-leaderboard}, and environment-setup benchmarks are already mined as RL training corpora \citep{savelov2025piper}, so held-out evaluation is needed. Existing benchmarks cover development-environment setup \citep{eliseeva2025envbench,arora2025setupbench,hu2025repo2run}, live operations \citep{shetty2025aiopslab,jha2026itbench}, and multi-stage DevOps workflows \citep{tang2026devopsgym}. FDE-Bench targets day~0 and day~1 deployment configuration: submitted artifacts are independently rebuilt and checked against a deployment specification.

We introduce FDE-Bench (Figure~\ref{fig:overview}): 136 tasks from a 149-card hand-authored blueprint corpus, with a four-arm release gate defining acceptance criteria. Tasks pair natural-language deployment tickets with \texttt{spec.yaml} declaring customer-observable requirements. Greenfield tasks require new deployment configurations; repair tasks start from broken ones. The tasks cover containerization, orchestration, health monitoring, and a full-pipeline family composing all three. Replay snapshots declarative artifacts, tears down the live environment, and re-applies them in a pristine sandbox for hidden checks. Four gated binary, programmatic layers prevent credit downstream of failed builds or crash-looping pods; live state earns no credit, and scoring uses no LLM judge.

A resolved episode means that the submitted artifacts satisfy the deployment specification on fresh replay under the task budgets. The evaluation asks whether the zero-intelligence strategies solve tasks (none resolves any evaluated task, with mean Deployment Score between 0.22 and 0.44 against a 100\% Resolved Rate for the reference solution), how much work agents complete (the best resolves 75\%), how the task groups compare (mean repair RR is 30.7 percentage points higher, with a positive gap for all seven models), and where deployments fail, which the check-layer and termination-status breakdowns localize.

We contribute the benchmark (136 released tasks from a 149-card blueprint corpus, with a public dev split, a held-out split under wave-based rotation, fault operators grounded in real incidents, and a per-task release gate), the protocol (replay grading through four gated binary programmatic layers, the DS and RR metrics, and an agent sandbox with allow-listed, logged registry egress), the evaluation (seven models using the same four-tool scaffold against three zero-intelligence floors and a reference skyline, as macro-averaged RR with Wilson intervals), and the validity analysis (complete failure-stage attribution and a diagnostic taxonomy, a spec-leak audit, gameability arms in the release gate, a sandbox-isolation audit, and contamination controls).

\section{Related Work}
\label{sec:related}

The appendix compares task scope and evaluation criteria across related benchmarks.

\paragraph{Development-environment setup.}
EnvBench \citep{eliseeva2025envbench} measures Python import errors and JVM compilation outcomes, and SetupBench \citep{arora2025setupbench} has agents install toolchains and resolve dependencies in bare sandboxes, validated by exit-zero commands. Installamatic \citep{milliken2025installamatic} assembles Dockerfiles from retrieved documentation, Repo2Run \citep{hu2025repo2run} builds images that run a repository's tests, and PIPer \citep{savelov2025piper} trains compact models for the same objective; Multi-Docker-Eval \citep{li2025multidockereval}, RAT \citep{zhang2026rat}, and ResearchEnvBench \citep{chen2026researchenvbench} extend the recipe across ecosystems and artifact types. A working developer environment is the central objective of this line of work. FDE-Bench instead makes replica conformance, connectivity, probe semantics, and alerting explicit deployment requirements.

\paragraph{Deployment and operations.}
DevOps-Gym \citep{tang2026devopsgym} evaluates build configuration, runtime monitoring, issue resolution, and test generation in Java and Go projects, including combined workflows. FDE-Bench evaluates Docker, Compose, and Kubernetes deployment artifacts for fixed applications through fresh replay. DeployBench \citep{wang2026deploybench} deploys research artifacts and grades a one-shot batch outcome. AIOpsLab \citep{shetty2025aiopslab} and ITBench \citep{jha2026itbench} evaluate day~2 operations on systems that arrive already deployed, so the configuration work FDE-Bench targets is furnished rather than tested. NetArena \citep{kim2026netarena} resists contamination by generating scenarios on demand; our held-out split instead rotates in waves drawn from the blueprint corpus's unreleased remainder. Infrastructure-as-code evaluation \citep{kon2024iaceval,gupta2025multiiacbench,mehta2026verifieriac} scores configuration text against schemas and policies, yet a well-formed manifest can describe a system that never becomes healthy; FDE-Bench grades the replayed deployment's build, readiness, behavior, and conformance instead.

\paragraph{Repository-level software and ML engineering.}
SWE-bench \citep{jimenez2024swebench} made issue resolution in a pre-configured repository the canonical agentic coding task, and MLE-bench \citep{chan2024mlebench}, MLAgentBench \citep{huang2024mlagentbench}, and RE-bench \citep{wijk2024rebench} are its ML and research-engineering analogues. These benchmarks center evaluation on code changes. In FDE-Bench, deployment configuration is the object of evaluation, application code is fixed, and checks run on the deployment reconstructed from submitted artifacts.

\paragraph{Terminal and OS generalists.}
Terminal-Bench \citep{tbench2026terminalbench}, OSWorld \citep{xie2024osworld}, and AgentBench \citep{liu2024agentbench} cover shell, desktop, and interactive tasks. SWE-agent \citep{yang2024sweagent} studies interfaces for repository navigation, editing, and execution; OpenHands \citep{wang2025openhands} integrates code, command-line, and browser interaction in a sandboxed platform. Our seven-model baselines hold the four-tool interface fixed; replay and gated checks measure deployment outcomes.


\section{The FDE-Bench Benchmark}
\label{sec:benchmark}

\begin{figure}[t]
\centering
\includegraphics[width=\textwidth]{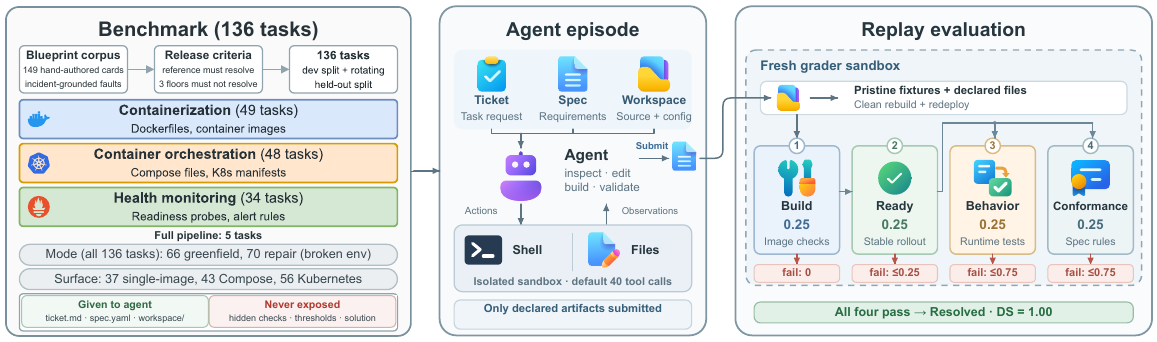}
\caption{FDE-Bench and its replay protocol. Left: task construction, composition, and agent-visible inputs. Center: the agent inspects, edits, builds, and validates the workspace before submitting declared artifacts. Right: a fresh grader rebuilds the submission over pristine fixtures. Build and readiness gate parallel behavior and conformance checks, each weighted 0.25. Red labels show score bounds after a failure; resolution requires all four layers.}
\label{fig:overview}
\end{figure}

FDE-Bench comprises 136 tasks from a 149-card hand-authored blueprint corpus; a four-arm release gate defines acceptance criteria. Agents build container images, author or repair orchestration manifests, and configure health monitoring for production-like deployments, submitting declarative artifacts for fresh re-deployment (Figure~\ref{fig:overview}).

\subsection{Task Definition}
\label{sec:tasks}

A task is a tuple $\mathcal{T} = (t, y, w, E, C, \rho)$. The natural-language deployment ticket $t$ uses a customer-facing engineer's voice. Specification $y$ (machine-readable \texttt{spec.yaml}) declares only customer-observable requirements: services and replicas, client-facing ports, endpoints including cross-service round-trips, health-endpoint semantics, monitoring, and global constraints such as no plaintext secrets. Workspace $w$ contains pinned-commit application source and partial or deliberately broken deployment configuration. $E$ pins the sandbox's container runtime, Kubernetes distribution, and component versions and declares the grading entrypoint. Hidden checks $C$ evaluate the deployment; reference solution $\rho$ checks solvability. In one shipped Compose task, $t$ requests an inventory API backed by a catalogue database and a Prometheus scrape; $y$ specifies published ports, a \texttt{/ready} endpoint returning 200 only while the catalogue is reachable, and one queryable metric series. The agent chooses base images, probe cadence, and other implementation details subject to remaining task constraints.

\paragraph{What the agent sees, and why.}
The agent observes $t$, $y$, and $w$ and interacts freely with $E$; $C$ and $\rho$ are never exposed. Rule disclosure states in the ticket that submissions are re-deployed fresh, making reproducibility explicit. Implementation concealment hides check probes, payloads, and thresholds, as with SWE-bench's hidden tests \citep{jimenez2024swebench}, to favor satisfying the specification over reverse-engineering the grader. The appendix includes a worked specification excerpt.

\paragraph{Budgets and terminal statuses.}
Each episode has a task-specific cap of 35--55 tool calls (40 for 114 of the 136 evaluated tasks), 20--35 wall-clock minutes, and 200{,}000 cumulative output tokens with at most 16{,}000 per turn. Caps are fixed across models for each task. A separate 2--3~USD kill-switch bounds API spending. The termination record distinguishes agent submission (\texttt{submitted}), step and time limits (\texttt{max\_steps}, \texttt{timeout}), output-token truncation (\texttt{max\_tokens\_truncated}), and cumulative output-token or cost limits (\texttt{budget}). Replay grades the resulting artifacts under the same deployment specification.

\subsection{Deployment Scenarios}
\label{sec:scenarios}

Tasks span three axes that mirror an FDE's day~0 and day~1 responsibilities, containerization (C), orchestration (O), and health monitoring (H), with per-axis, per-mode, and per-surface counts in the benchmark card and composition table in the appendix. A full-pipeline family (F) requires all three axes in one episode. The tasks form 14 scenario families, with per-family composition in the appendix; at $n=5$ the F family is reported as a case study and excluded from per-axis aggregates.

\paragraph{Greenfield and repair.}
Greenfield tasks are authored from source and specification alone; repair tasks hold a complete but deliberately broken environment to diagnose and fix. Broken environments are built by golden-config mutation: fault operators grounded in real community incidents are applied to a validated reference configuration, each recorded with the check layer it kills. A mutant is admitted only if it breaks at least one required check, so every repair task has an observable failure rather than a cosmetic diff.

\paragraph{Task sources.}
Every task workspace is an original application authored for the benchmark. Its services imitate common production shapes such as web APIs, workers, caches, and databases, with health-semantics quirks such as slow warm-up, lazy database connection, and gradual memory ramp that defeat copy-pasted probe defaults. They derive from no existing codebase, so no public ancestor exists to memorize; realism comes from provenance, blueprint cards grounded in real community incidents on Stack Overflow, GitHub issues, and Kubernetes forums, each recording its sources.

\paragraph{Construction quality assurance.}
Each released task must pass a four-arm mechanical release gate. The hidden reference solution must resolve ($s=1.0$). Three other arms must all fail to resolve: a do-nothing arm submitting the workspace unmodified, a spec-transcription arm reading only the agent-visible specification, and a generic stub arm shipping a minimal standard-library listener. An arm that errors counts as a gate error, never a pass. These criteria test solvability and reject tasks solved by the three specified baseline strategies; a single gate run does not establish grading stability. Tasks also include \texttt{solution\_b}, a second implementation held to the same must-resolve criterion. The frozen evaluation records contain successful second-solution checks for 132 of the 136 tasks, and the second solution also resolves each of the 64 tasks re-gated for the v1.1 revision (Section~\ref{sec:setup}), providing a check against dependence on one implementation.

\paragraph{Contamination design.}
Every task's files embed a canary GUID following the BIG-bench convention \citep{srivastava2022bigbench} for post-hoc detection in training corpora. The sandbox's egress proxy logs every request as a per-episode audit trail. Roughly 30 tasks form a public dev split shipping checks and solutions. The remainder is a held-out test split shipping only the agent-visible surface, graded through maintainer-run submissions. Retired waves move to the dev split, replacements are materialized from the unreleased remainder of the blueprint corpus, task identity is tracked by card id, and results are never compared across waves. Because every application is original, memorization pressure concentrates on the released text itself, which the canaries monitor.

\subsection{Sandbox Evaluation Protocol}
\label{sec:protocol}

\paragraph{Isolation and reproducibility.}
Each episode gets a Docker-in-Docker daemon and an agent shell container on an internal network, torn down afterwards. Agent egress passes through a logged proxy allowing image, OS, and language package registries. Grader-side builds had unrestricted egress. The server sweep used 12 single-node kind clusters, with at most two grading episodes per cluster. Each episode has a separate namespace and ephemeral host ports for probes. The grid also retains earlier-host episodes; the appendix records the execution mix and settings. Live probes check isolation. The privileged daemon targets accidental and opportunistic access. Base images are version-pinned, and reference re-verification detects registry drift before evaluation.

\paragraph{Scaffold.}
Agents use four tools: \texttt{bash}, a persistent shell with a 600\,s per-command cap and truncated output, plus \texttt{write\_file}, \texttt{read\_file}, and \texttt{submit}. A fixed system prompt, identical across models except for the task's budget line, controls prompt and tool differences. Models must support native function calling; after one prose reminder, episodes that stop calling tools terminate as \texttt{no\_tool\_use}.

\paragraph{Replay grading.}
At submission, the harness (1)~collects only files matched by the task's declared \texttt{graded\_artifacts} globs, such as Dockerfiles, manifests, and monitoring configuration; (2)~tears down the live environment entirely; (3)~overlays the files onto pristine fixtures and re-applies them through the task's entrypoint in a fresh grader sandbox; and (4)~runs the hidden check suite. Let $w'$ denote the submitted workspace, $\Pi_g$ the projection onto the graded globs, and $F$ the pristine fixtures. The episode score is
\begin{equation}
\label{eq:replay}
s(w') \;=\; \Phi\bigl(C,\ \mathrm{Deploy}_{E}(\Pi_g(w'),\,F)\bigr).
\end{equation}
By construction $s(w') = s(\Pi_g(w'))$. Imperative live state, such as a hand-started container or an injected pod secret, is not in the image of $\Pi_g$ and earns nothing. Edits outside the graded globs, application source included, are discarded and rebuilt from scratch.

\paragraph{Gated check layers.}
Checks form four gated layers. L1 build: artifacts build and respect declared image constraints. L2 readiness: the rollout completes and workloads reach and hold Ready across task-defined stability windows. L3 behavior: endpoints answer seeded probes with correct cross-service round-trips, required PromQL series are present, and alerts fire under injected fault yet stay silent through a healthy control window. L4 spec conformance: the deployment respects the replica counts declared in the specification, image-size quotas, non-root execution, and the ban on plaintext secrets.

\paragraph{Scoring.}
This revision assigns each layer weight 0.25 and recomputes DS from the retained binary verdicts. The per-episode score is
\begin{equation}
\label{eq:score}
s = 0.25\,(b + b\,r + b\,r\,e + b\,r\,c),
\end{equation}
where $b, r, e, c \in \{0,1\}$ indicate that each layer passes in full. Build and readiness gate both behavior and conformance; the latter two are parallel requirements. Binary layers give each task the same score lattice, $\{0, 0.25, 0.50, 0.75, 1.00\}$, independent of its number of checks. An upstream failure blocks downstream functional assessment, so DS measures achieved deployment stages rather than the fraction of correctly authored configuration. For example, passing conformance while failing behavior yields $s=0.75$. An episode is Resolved iff all four layers pass ($s=1.0$). RR is the fraction of tasks resolved, reported with 95\% Wilson intervals; mean DS and the failure-stage breakdown provide secondary diagnostics.

\paragraph{Flakiness engineering.}
Deployment checks interact with schedulers, image caches, and network timing, so every timing assertion is a stability window rather than an instantaneous timeout: restart counts must hold over a window, not at an instant.

\paragraph{Why no LLM judge.}
All scoring uses programmatic checks; no model interprets submitted manifests or logs as grading instructions. The pinned harness and task fingerprints support replay and auditability, while scheduler and network timing remain sources of variation. This approach requires specifications that can be evaluated programmatically.


\section{Experiments}
\label{sec:experiments}

\subsection{Setup}
\label{sec:setup}

\paragraph{Model lineup.}
We evaluate seven language models from four independent providers: Gemini-3.6-Flash and Gemini-3.5-Flash-Lite, GPT-5.6-Terra and GPT-5.6-Luna, Claude-Sonnet-5, and the open-weight pair DeepSeek-V4-Pro and DeepSeek-V4-Flash. Each uses its provider's first-party API; recorded SDK versions and request settings are listed in the appendix. Models use native function calling.

\paragraph{Evaluation grid.}
We report the v1.1 grid: all 136 released tasks, 66 greenfield and 70 repair, with no exclusions. Relative to the frozen v1.0 grid, 29 tasks now state in \texttt{spec.yaml} the stability window their checks already graded, and the shared Kubernetes workload-discovery check now starts from the port or in-cluster URL the contract names and accepts several owning workloads. Three Kubernetes tasks return after their packaged-reference and check-suite defects were fixed. All 64 affected tasks, the 29 with revised specifications and 35 Kubernetes tasks, pass the release gate again. Every model was re-run once on the 32 revised or returning tasks (224 episodes), and the Gemini-3.6-Flash \task{export-jobproc-metrics} episode, invalid in v1.0, was re-run, giving 225 fresh episodes at seed 900 beside 727 frozen v1.0 cells; one fresh API-error episode (Gemini-3.6-Flash on an image repair task) was re-rolled once, and the appendix section on disclosed task defects and grid membership records the changes.

\paragraph{Floor baselines.}
Three zero-intelligence floors, the release-gate arms run on every task, bound the leaderboard from below. The do-nothing arm submits the workspace unmodified. The spec-transcription arm mechanically transcribes the agent-visible specification into artifacts without opening the source, a canary for specifications that publish their own answers. The generic stub arm ships a minimal standard-library listener on the declared port, a canary for checks that only prove something answered. A fourth floor, a rule-based template heuristic emitting stock Dockerfiles and generic manifests, tests whether the tasks resist boilerplate.

\paragraph{Scaffold policy.}
All seven models use the same four-tool loop and fixed prompt, with zero per-model tuning. Reported success rates, step counts, and failure patterns characterize each model under this shared interface and the task budgets.

\paragraph{Episodes and metrics.}
Each (task, model) cell has one episode with harness seed 900 under the per-task budgets. Temperature, top-p, penalties, and API sampling seed were omitted from requests, leaving provider defaults in effect. The protocol forbids re-rolling to select a more favorable outcome. Graded episodes are cached for auditing. The cache checks task and agent-surface fingerprints when present; legacy records without these hashes require separate version verification. Stability windows reduce sensitivity to transient states, but this single-episode evaluation does not estimate variation across repeated runs.

\paragraph{Statistical methodology.}
Confidence intervals on RR are 95\% Wilson intervals \citep{wilson1927} over the per-task binary outcomes. \texttt{harness\_error} and \texttt{api\_error} episodes are excluded from every rate, and an agent's evaluation is invalid if such episodes exceed 10\% of its episodes. Table~\ref{tab:main} reports recorded step/time-cap counts; the appendix separates output-token truncations and unknown termination statuses. Pairwise comparisons use family-clustered sign-flip permutation tests on per-task resolved-outcome differences over the 14 families, because within-family variants share structure; the statistic is given in the appendix.

Two reference comparisons complement the floors: the reference-solution oracle replays the author solutions on the evaluation grid, where they resolve all 136 tasks, and the ensemble union counts a task resolved if any evaluated agent resolves it.

\subsection{Main Results}
\label{sec:results}

Table~\ref{tab:main} reports overall RR and DS, mean DS by axis and mode, cap counts, and median steps. The appendix also reports RR and resolved/valid episode counts by axis and mode.

\begin{table}[t]
\caption{Main results on the 136-task grid, one cached episode per cell, with 95\% Wilson intervals on RR. DS columns report mean Deployment Score; Caps lists recorded step/time terminations as total/unresolved. The full-pipeline family ($n=5$) is excluded from axis columns and reported separately in the appendix. Every model has 136 valid episodes. The template covers 109 tasks. The reference skyline resolves every task.}
\label{tab:main}
\centering
\footnotesize
\setlength{\tabcolsep}{2.5pt}
\begin{tabular}{@{}l ccc ccc cc c@{}}
\toprule
& \multicolumn{3}{c}{Overall} & \multicolumn{3}{c}{DS by axis} & \multicolumn{2}{c}{DS by mode} & \\
\cmidrule(lr){2-4} \cmidrule(lr){5-7} \cmidrule(lr){8-9}
Agent & RR \% (95\% CI) & DS & Caps & C & O & H & Greenfield & Repair & Med.\ steps \\
\midrule
Gemini-3.6-Flash & 75.0 (67.1--81.5) & 0.857 & 42/15 & 0.878 & 0.807 & 0.919 & 0.777 & 0.932 & 35 \\
GPT-5.6-Terra & 69.9 (61.7--76.9) & 0.842 & 2/0 & 0.883 & 0.766 & 0.919 & 0.780 & 0.900 & 9 \\
Claude-Sonnet-5 & 72.8 (64.8--79.6) & 0.855 & 24/9 & 0.918 & 0.828 & 0.853 & 0.761 & 0.943 & 29 \\
DeepSeek-V4-Pro & 69.9 (61.7--76.9) & 0.831 & 53/22 & 0.791 & 0.839 & 0.897 & 0.723 & 0.932 & 35 \\
GPT-5.6-Luna & 63.2 (54.9--70.9) & 0.785 & 0/0 & 0.786 & 0.771 & 0.846 & 0.720 & 0.846 & 11 \\
DeepSeek-V4-Flash & 66.2 (57.9--73.6) & 0.811 & 44/18 & 0.837 & 0.776 & 0.831 & 0.682 & 0.932 & 30 \\
Gemini-3.5-Flash-Lite & 52.9 (44.6--61.1) & 0.719 & 61/39 & 0.643 & 0.776 & 0.779 & 0.576 & 0.854 & 37.5 \\
Ensemble union           & 92.6 (87.0--96.0) & -- & -- & -- & -- & -- & -- & -- & -- \\
\midrule
Rule-based template & 0.0 (0.0--3.4) & 0.275 & -- & 0.203 & 0.540 & 0.205 & 0.236 & 0.311 & -- \\
Generic stub & 0.0 (0.0--2.7) & 0.438 & -- & 0.480 & 0.469 & 0.360 & 0.439 & 0.436 & -- \\
Spec-transcription & 0.0 (0.0--2.7) & 0.335 & -- & 0.276 & 0.422 & 0.338 & 0.269 & 0.396 & -- \\
Do-nothing & 0.0 (0.0--2.7) & 0.217 & -- & 0.168 & 0.203 & 0.338 & 0.004 & 0.418 & -- \\
\midrule
Reference skyline & 100.0 (97.3--100.0) & 1.000 & -- & 1.000 & 1.000 & 1.000 & 1.000 & 1.000 & -- \\
\bottomrule
\end{tabular}
\end{table}

\paragraph{Overall ordering and the floor structure.}
Gemini-3.6-Flash has the highest observed RR, 75.0 percent, and the highest DS, 0.857; Claude-Sonnet-5 is second on both, at 72.8 percent and 0.855. The pairwise tests support some comparisons but not the full ordering: the lead of Gemini-3.6-Flash over its Flash-Lite sibling is supported ($p=0.0015$ under the family-clustered permutation test), its lead over Claude-Sonnet-5 is not ($p=0.59$), and GPT-5.6-Terra and DeepSeek-V4-Pro resolve identical task counts. The floors behave as designed: none resolves a single task, and their DS means sit below the lattice midpoint, with the floor and reference score-lattice distributions in the appendix. The do-nothing floor also exposes an asymmetry to read before interpreting any mode gap: repair workspaces start partially passing by construction, at 0.418 mean DS against 0.004 for greenfield, so raw DS on repair embeds a floor that RR does not.

\paragraph{Per-axis pattern.}
We hypothesized that health monitoring is the hardest single axis, because an H behavior check requires the alert to fire while the fault is injected and to stay silent on the control run, so a plausible but unwired rule fails one half or the other. The data refute the hypothesis. For the best agent the axis profile is C 73.5, O 72.9, and H 82.4 percent RR, and six of seven agents resolve H at or above their C rate. Macro-averaged RR is lowest on orchestration, 65.2 percent, against 67.1 on containerization and 73.1 on health monitoring. The ten tasks no agent resolves are all greenfield and concentrate in containerization: seven are single-image C tasks, where a wrong base image, dependency pin, or image-hygiene constraint fails in ways that survive forty steps of iteration, and the other three are Kubernetes tasks. The conjunctive full-pipeline family is reported as a case study in the appendix: four of its five tasks resolve for at least one agent, none for all seven, and every zero-intelligence arm floors out across all five. Across the two task groups, macro-averaged RR is 82.0 percent for repair and 51.3 percent for greenfield, a gap of 30.7 percentage points with the same sign for all seven models.

\paragraph{Calibration and skylines.}
Difficulty calibration targeted a best-agent RR of 35--60 percent at release, and the strongest agent exceeds that band at 75.0. The ensemble union resolves 92.6 percent, 17.6 percentage points above the best single agent, showing that different agents solve different tasks. Reference solutions resolve all 136 tasks. Ten tasks remain unresolved by all seven agents. A separate Claude-Opus-5 probe on the historical task versions resolved four of the ten, two single-image and two Kubernetes tasks, leaving six unresolved across the recorded model runs.

\section{Analysis}
\label{sec:analysis}

The analyses use the retained layer verdicts; equal weighting changes partial DS while preserving resolution and failure-stage assignments.

\subsection{Failure-stage breakdown}
\label{sec:waterfall}

\begin{figure}[t]
\centering
\includegraphics[width=\textwidth]{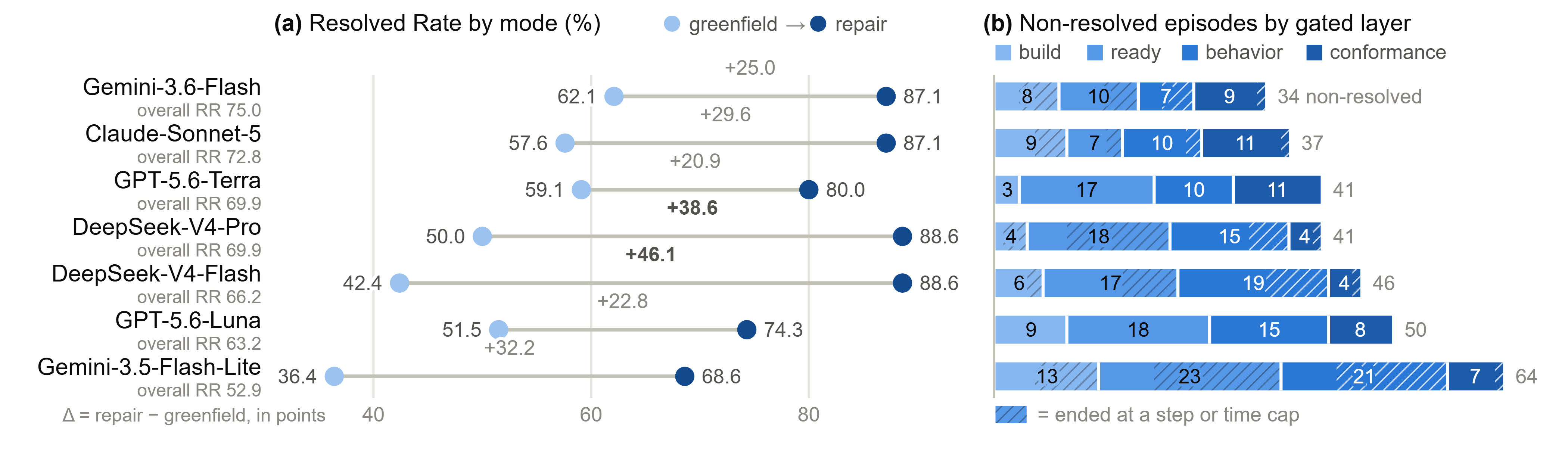}
\caption{Results on the 136-task grid, models sorted by overall RR. (a)~Every evaluated model has a higher RR on repair tasks than on greenfield tasks; the two DeepSeek models have the largest gaps. The grid contains disjoint groups of 70 repair and 66 greenfield tasks, with one episode per task--agent cell. (b)~Non-resolved episodes stacked by the gated layer they failed, with hatching for episodes ended by a step or wall-clock cap; fail-at-ready is the largest pooled state, and the mix is model-specific.}
\label{fig:results}
\end{figure}

We assign each episode to one failure stage in the order build, readiness, behavior, conformance, or resolved, at $s=0$, $0.25$, $\{0.50,0.75\}$, $0.75$, and $1.0$, respectively. This reporting priority makes the categories disjoint: behavior-fail includes episodes that pass conformance. The score retains the parallel behavior and conformance branches. An orthogonal cap-limited flag marks episodes ended by a step or wall-clock limit; capped episodes are graded through the same gates and land in the same five states. Figure~\ref{fig:results}b shows the per-model failure mix.

Hypothesis H1 predicted that the modal way to fail is the ``it builds but does not run'' outcome: Dockerfile and manifest authoring are static text generation, well represented in public corpora, whereas readiness requires closed-loop interaction with a live cluster. The data confirm the modal claim while bounding its size: 12.2 percent of build-passing episodes fail to reach a stable Ready rollout, making fail-at-ready the largest failure state, 110 of 313, though the mix is model-specific, with Claude-Sonnet-5 failing most at conformance and DeepSeek-V4-Flash at behavior. The gated checks distinguish failed builds and unstable rollouts from behavior and conformance failures.

\subsection{Greenfield versus repair}
\label{sec:greenfield-repair}

The evaluation uses disjoint sets of 70 repair tasks and 66 greenfield tasks that share scenario families and specification style. Repair tasks start from deliberately broken deployment configurations. Figure~\ref{fig:results}a shows a higher repair RR for every evaluated model, with gaps from 20.9 percentage points for GPT-5.6-Terra to 46.1 percentage points for DeepSeek-V4-Flash. Mode-specific rankings expose differences hidden by overall RR: DeepSeek-V4-Flash ranks sixth on greenfield and ties DeepSeek-V4-Pro for first on repair, while GPT-5.6-Terra ranks second and fifth, respectively.


\paragraph{Step efficiency.}
The median step count distinguishes the evaluated agents more sharply than RR: GPT-5.6-Terra and Claude-Sonnet-5 resolve similar numbers of tasks, 95 and 99, at medians of 9 and 29 tool calls, and DeepSeek-V4-Pro and DeepSeek-V4-Flash spend most of a typical 40-call budget at medians of 35 and 30. Four models reached a step or wall-clock limit in 42 to 61 episodes. GPT-5.6-Luna recorded no such terminations and still trails DeepSeek-V4-Pro; two Luna episodes ended with output-token truncation.

\subsection{Failure taxonomy and validity}
\label{sec:taxonomy}

Replay checks localize all 313 unresolved episodes to a failure stage: 52 at build, 110 at readiness, 97 at behavior, and 54 at conformance. Step or wall-clock caps ended 103 of these episodes (32.9\%), including 38 whose artifacts passed both build and readiness. Of the 54 conformance failures, 49 followed an agent submission and five ended at a cap. The stage-by-termination table in the appendix links these counts to the recorded outcomes; its diagnostic taxonomy supplies categories for examining checks and logs. One Claude-Sonnet-5 failure matches the task disclosure defect described below and remains counted in Table~\ref{tab:main}.

Three validity audits are already reflected in the released benchmark. First, a spec-leak audit found 58 of 61 hidden check constants restated verbatim in early specifications, and a zero-intelligence transcriber resolved the pilot tasks outright, voiding those numbers. Every agent-visible file was scrubbed to customer-observable obligations, and the spec-transcription and generic stub arms became permanent gate arms. Second, a sandbox-isolation audit found an early configuration exposing the grader's Docker socket to agent-side processes; it was replaced, and isolation is now asserted by live probes. Third, contamination controls consist of corpus-wide canary GUIDs \citep{srivastava2022bigbench} and the egress proxy's per-episode access log. Stability windows reduce sensitivity to transient conditions but do not quantify flakiness. Repeated replay at scale remains future work.

\paragraph{Difficulty headroom.}
On the initial twelve-task probe set, Claude-Opus-5 resolved eleven tasks in 5 to 46 steps without a cap. These July 28, 2026 probes used the historical task versions. The sole failure, \task{instrument-ledger-api-busy-metrics}, exposed a task disclosure defect: replay successfully built the submitted image, then failed while pulling its local tag from a registry. The task's verification text described build and deploy but omitted this pull stage.

Six additional tasks combined replay-hostile configurations, conjunctive requirements, timing conditions, and tight budgets. Four have probe records: three tasks resolved in valid episodes; \task{deploy-https-storefront-with-bundle} ended with an API error and yielded artifacts scoring 1.0 on replay. We retain it as an artifact-level diagnostic and exclude it from the valid-episode count. The remaining two, both Kubernetes tasks, have no recorded probe. One of the three left the frozen grid after check-suite crashes and returns in v1.1, where no roster model resolves it. The probe ledger in the appendix lists all tasks, validity decisions, and recorded model settings.

\paragraph{Human baseline.}
One practicing junior FDE, who was not involved in benchmark construction and had no access to grading assets, completed a stratified 25-task subset serially in one pass. The participant directed Claude Code with Claude-Sonnet-5 fixed as the model and web search disabled. Submissions were replay-graded through the unmodified pipeline. The human resolved 23 of 25, against 18 for the autonomous Claude-Sonnet-5 episodes on the same tasks (17 on the frozen grid, which excluded one subset task), winning six of the seven discordant cells at $p=0.125$ under a sign test \citep{mcnemar1947}. The participant resolved \task{containerize-feedback-api-read-only}, which no roster agent resolved; both failures in the 25-task subset occurred at conformance. In a separate hard-probe task outside that subset, the participant also resolved \task{instrument-ledger-api-busy-metrics}, where the disclosure defect had caused both Anthropic models' autonomous episodes to fail. Three limitations qualify this comparison: one participant, session timing was not logged, and the comparison confounds human judgment with the richer scaffold.

\section{Limitations and Conclusion}
\label{sec:limitations}

Each model--task cell contains one episode, so Wilson intervals describe variation across tasks. The reference replay archive contains matching check outcomes on 30 development tasks. All 30 have two runs with matching task fingerprints; seven have a third, giving 67 runs and 1{,}138 check observations. Complete oracle-artifact and shared-grader identities were not retained. The appendix records this coverage; timed checks remain sensitive to host load.

The adversarial evaluation tests three author-designed attacks against the corresponding grading checks. Paired weakened verifiers test whether each payload activates its intended shortcut. Broader attack coverage requires independently designed submissions.

The released rubric records seven graded behaviors without a stated specification clause, on three Kubernetes tasks and one Compose task; v1.1 closed the other 56 of the 63 declared in v1.0 by stating the stability window in 29 specifications and by anchoring Kubernetes workload discovery on the contract. Two grid episodes, DeepSeek-V4-Pro and DeepSeek-V4-Flash on \task{deploy-corridor-api-no-outage-k8s}, fail two of the seven checks and also fail 11 checks that a specification clause states, so no reported resolution or Deployment Score turns on a gap; the other five checks fail no episode. This bound matches episodes to the current rubric by task name rather than by task version. Repairing the remaining contracts requires a new task version and renewed evaluation. The human comparison covers one participant using a richer interface. Specialized agent interfaces may change the model results under the same task and grading protocol.

\label{sec:conclusion}
FDE-Bench makes the deployed environment the object of evaluation. Under a shared four-tool scaffold, the strongest evaluated model resolves 75.0 percent of tasks, and ten tasks remain unresolved by all seven models. Readiness accounts for 110 of 313 failures; the disjoint repair task group has a mean RR 30.7 percentage points higher than greenfield. Fresh replay and check-layer diagnostics connect these outcomes to reproducible deployment artifacts. Comparisons across agent interfaces and repeated runs can build directly on this evaluation protocol.

\clearpage
\section*{Reproducibility Statement}

The historical task and model records are archived under v1.0.0-rc1 with an MIT license. The repository contains 136 tasks, the harness and grader, release-gate verdicts, a 220-check offline probe suite with live isolation probes, and protocol documentation. Gate evidence applies to its archived task version. The v1.1 revision keeps the same 136 tasks and every v1.0 record, and adds the 29 revised specifications, the corrected workload-discovery rule, the release-gate ledgers for the 64 re-gated tasks, and the 225 fresh episodes.

This revision recomputes DS from retained binary verdicts with equal layer weights, identified as \texttt{fde-ds-equal-v2}. Original scores use 0.20/0.25/0.35/0.20 and remain archived; both versions require all four layers for resolution. Each model--task cell has one episode with harness seed 900. Cached records preserve the original outcomes, and the protocol forbids selecting a better score by rerunning. Legacy records without complete fingerprints require separate version verification.

The server sweep used a host recorded as 192 cores, 384 threads, and 503 GB RAM alongside other workloads; the grid includes earlier 16-thread laptop episodes. The appendix records SDKs, sampling settings, cluster isolation, and missing host metadata. Two release-gate arms failed under memory pressure and passed on an idle host, motivating a dedicated idle host for timed checks.

A task leaves the grid under one rule, that its instance yields no valid verdict for any model. Under v1.0 this removed \task{deploy-engineering-index-k8s}, whose packaged reference manifest had lost the foreground flag its server needs, and \task{release-token-svc-without-drops-k8s} and \task{autoscale-payments-api-k8s}, whose check suite raised \texttt{TypeError} after an event named \texttt{\_stop} shadowed the thread method; v1.1 fixes both defects and returns all three tasks. A ticket that under-describes the grading procedure does not meet that rule. Task \task{instrument-ledger-api-busy-metrics} omits a pull stage from its verification text, yet its reference resolves and six of seven models clear the build layer, three of them resolving the task, so it stays in the grid; the appendix reports the per-model effect of removing it. New task and grader versions retain separate results. The appendix reports three current-version \task{containerize-metrics-shim} reference replays with complete task, grader, and artifact hashes; all 13 checks agree across the three deployments. Subsequent Harbor parity tests have their own task, artifact, and execution records.

\bibliographystyle{iclr2027_conference}
\bibliography{references}

\section*{AI Use Statement}

We used generative AI tools to refine research ideas, the benchmark methodology, and experimental designs. These tools assisted with implementing and debugging research code, cleaning and reformatting experimental logs, and analyzing and interpreting results. They also supported literature retrieval and summarization, manuscript drafting and restructuring, language editing, figure preparation, and document and reference formatting. The evaluated language models generated the agent actions and deployment artifacts studied in the experiments. The replay evaluator computed scores using programmatic checks, without an LLM judge.

The authors reviewed and revised the AI-assisted text and checked the research claims against the reported methods and evidence. We inspected and ran the experimental and analysis code, checked results against retained configurations and logs, verified figures against the underlying results, and checked cited references against their sources. We also compiled and inspected the manuscript. The authors take responsibility for the final text, claims, code, analyses, and artifacts, including content produced with generative AI assistance.

\clearpage
\appendix
\section{Worked Task Example}
\label{app:example}

This worked example shows how a specification fragment defines the deployment obligations.

\paragraph{Worked example.}
The following is condensed from the shipped specification of a three-service task in the full-pipeline family (Compose surface). Every specification opens with a preamble listing deliberately omitted implementation choices: base images, process models, schema-application timing, credential delivery, scrape configuration, probe timings, and all PromQL. These choices are left to the agent. \texttt{contract\_numbers} lists the only constants a hidden assertion may share with the agent-visible text.

{\scriptsize
\begin{verbatim}
stack: depot-inventory
services:
  api:        # the inventory service the
              # yard's tablets call
    port_published: "18400:8000"
  db:         # the parts catalogue, answering
              # inside the stack on 5432
  prometheus: # collects the api's request counts
    port_published: "18132:9090"
contract:
  api:
    reachable_at: host port 18400
    endpoints:
      - GET  /ready -> 200, and 503 while the
        catalogue cannot be served
      - POST /items -> 201 the stored part;
        400 when the part is not described
    write_then_read: a part added through the
      api is readable afterwards, and is
      really in the catalogue
  catalogue:
    survives_container_replacement: a part
      added before the replacement is still
      readable after it
    requires_the_issued_credential: a
      connection with the wrong password
      is refused
  monitoring:
    queryable: inventory_http_requests_total
\end{verbatim}
}

\noindent To resolve this contract, the agent must build both images from source during the session and publish the declared ports. It must apply the schema before traffic arrives, deliver the issued credential without plaintext, and provide catalogue storage that survives container replacement. It must also configure scraping so the depot's dashboards can query the counter. Everything the contract does not state (base image, server, worker count, probe cadence) remains the agent's engineering choice. The hidden checks generate requests at grading time to test behavior beyond fixed responses.

\section{Benchmark Card}
\label{app:card}

Following the documentation practice of \citet{gebru2021datasheets}, we summarize FDE-Bench in card form. Table~\ref{tab:card} gives the at-a-glance facts; the paragraphs below cover purpose, scope, and stewardship.

\begin{table}[t]
\centering
\footnotesize
\setlength{\tabcolsep}{3pt}
\caption{FDE-Bench at a glance.}
\label{tab:card}
\begin{tabular}{@{}ll@{}}
\toprule
Tasks / blueprint cards & 136 / 149 \\
Split (dev / held-out) & $\sim$30 / $\sim$106 \\
Axes (C/O/H/F) & 49 / 48 / 34 / 5 \\
Surfaces (image/Compose/k8s) & 37 / 43 / 56 \\
Greenfield / repair & 66 / 70 \\
Layer weights (gated) & .25/.25/.25/.25 \\
Metrics & RR (Wilson 95\% CI), mean DS \\
License & MIT \\
Task corpus / harness / score & v1.1 / 1.0.0-rc1 / equal-v2 \\
\bottomrule
\end{tabular}
\end{table}

\paragraph{Motivation and evaluative purpose.}
FDE-Bench measures whether an LLM agent can use a deployment specification and a target environment to produce container images, orchestration manifests, and health monitoring that satisfy the specification. Scores measure this ability within an isolated sandbox spanning three surfaces: single Docker images, Compose stacks, and single-node Kubernetes. The reported baselines evaluate seven models with the same four-tool scaffold and fixed prompt, under allow-listed registry egress and fixed step, time, and output-token budgets, with no per-model tuning. All checks are programmatic; no LLM judge contributes to any score.

\paragraph{Intended uses.}
Comparing agent systems on deployment-configuration competence; tracking progress across model generations under a frozen harness; failure analysis through check outcomes and diagnostic categories; and analysis of performance across repair and greenfield task groups, using the dev split for development and the held-out split for confirmatory reporting.

\paragraph{Out-of-scope uses.}
FDE-Bench is not a production-security certification: L1/L4 hardening checks verify specific properties, not the absence of vulnerabilities. Scores do not transfer to live multi-tenant clusters with real traffic, real credentials, or organizational change-management processes. The benchmark does not cover cloud-provider infrastructure-as-code (Terraform, CloudFormation), which is evaluated elsewhere \citep{kon2024iaceval, gupta2025multiiacbench}.

\paragraph{Contamination policy.}
Every task's agent-visible and withheld metadata files carry a canary GUID following the BIG-bench convention \citep{srivastava2022bigbench}, and the sandbox's egress-proxy access log records everything an agent fetched during a run. Held-out tasks ship only their agent-visible surface; retired waves move to the dev split and replacement waves are materialized from the unreleased remainder of the blueprint corpus, with task identity tracked by card id and results never compared across waves. Suspected contamination can be reported to the maintainers; confirmed reports trigger task retirement and a versioned leaderboard annotation.

\paragraph{Versioning policy.}
Releases follow semantic versioning: patch versions fix task bugs without changing the task set, minor versions add or retire tasks, and major versions change scoring. Benchmark-version and task-fingerprint metadata identify the conditions under which a leaderboard cell was graded, where those fields are recorded. Graded episodes are cached, but legacy records without fingerprints are retained and cannot by themselves establish equivalence to current task files. The protocol permits re-running a cell when its task's content changes and prohibits silently rescoring it.

\paragraph{Safety considerations.}
Agents execute arbitrary shell commands. Agent execution uses per-run Docker-in-Docker sandboxes on internal networks; egress passes through a logged proxy allowlisting image, OS, and language package registries. Run state is destroyed afterward. Historical grader-side builds had unrestricted egress. The Docker-in-Docker daemon runs privileged, so the configuration targets accidental and opportunistic access, not a hostile-model kernel escape; isolation is asserted by live probes run against every release. No real credentials, tokens, or external services appear anywhere in the benchmark. Users must not point the harness at shared or production clusters.

\paragraph{Known limitations.}
Tasks are confined to the Linux/x86 Docker-and-Kubernetes ecosystem and English-language tickets, and the original applications imitate popular web-service stacks. Checkers verify specification conformance without assessing configuration style. Task base images are version-pinned but not digest-pinned, so registry drift can alter difficulty between releases. The reference arm is re-run before evaluation to detect such changes. Release-gate outcomes do not estimate flake rates; large-scale repeat-replay certification remains future work.

\paragraph{Maintenance commitment.}
The maintainers commit to three years of support: task-bug reports triaged within 30 days, an annual task refresh alongside private-split rotation, and archival deposit of all public artifacts with a persistent identifier.

\section{Task Schema and Worked Example}
\label{app:schema}

Each task is a self-contained directory:

{\scriptsize
\begin{verbatim}
<task-id>/
  task.yaml      # manifest the harness consumes
  ticket.md      # natural-language ticket
  spec.yaml      # observable deployment contract
  workspace/     # agent starting state
  checks/        # HIDDEN: check suite
  solution/      # HIDDEN: reference solution
  solution_b/    # HIDDEN: second solution
\end{verbatim}
}

\texttt{task.yaml} is the manifest the harness consumes. Its identity fields include \texttt{card\_id}, the blueprint-card key that stays stable across waves, and provenance. The fields \texttt{id}, \texttt{family}, \texttt{mode}, and \texttt{kind} identify the task, its family for clustered comparisons, its greenfield or repair mode, and its image, Compose, or Kubernetes surface.

The grading contract specifies \texttt{graded\_artifacts}, the globs collected at submission; nothing else survives to replay. It also specifies \texttt{entrypoint}, the declared build and deploy commands executed during replay, and \texttt{host\_port} / \texttt{container\_port}. The \texttt{caps} field carries the budgets \texttt{max\_steps}, \texttt{max\_minutes}, and \texttt{max\_cost\_usd}. A corpus-wide canary header marks every file.

\texttt{ticket.md} is the natural-language deployment ticket, written as a realistic request from a client engineering team. It discloses that submitted artifacts will be re-deployed in a fresh environment for verification. \texttt{spec.yaml} states the observable contract (Appendix~\ref{app:example}): services and published ports, endpoint behavior including data round-trips, health semantics, monitoring requirements, and constraints. Its \texttt{contract\_numbers} field lists the only constants hidden assertions may share with agent-visible text, and its preamble lists deliberately omitted implementation choices. \texttt{workspace/} is the mutable starting state: application source plus partial or deliberately broken configuration, depending on mode.

\texttt{checks/} contains the hidden layer-check implementations executed only in the grader sandbox after replay. \texttt{solution/} contains the hidden author solution, which must resolve in the release gate. \texttt{solution\_b/} is a mechanically different second solution held to the same must-resolve requirement; the frozen evaluation records document successful runs on 132 of the 136 released tasks, and the v1.1 release gate re-ran it on 64 tasks (the 29 with changed specifications and 35 Kubernetes tasks), where it resolved every one.

As a worked example, one O2 repair task ships a workspace whose orchestration references a credential key renamed by the recorded fault operator. The ticket reports that the API pods never become ready since the last configuration change, and the contract requires the service ready and answering on its published port. The hidden checks verify rollout completion, endpoint behavior with requests minted at grading time, and that the fix restores the credential wiring rather than inlining a plaintext value.

\section{Corpus Composition}
\label{app:composition}

\begin{figure}[t]
\centering
\includegraphics[width=\columnwidth]{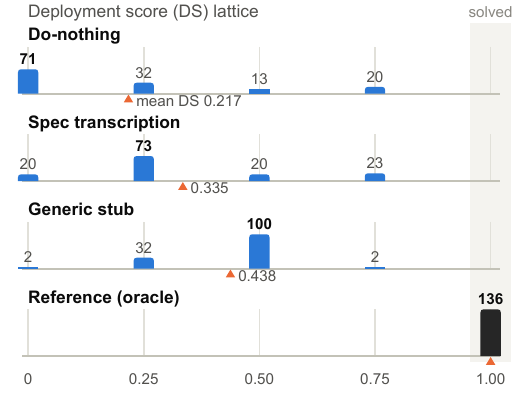}
\caption{Score distributions of the four release-gate arms over the 136-task v1.1 grid on the discrete score lattice. Each floor has a distinct modal score, do-nothing at $0$ on 71 of 136 tasks, spec-transcription at $0.25$ on 73, and the generic stub at $0.50$ on 100, and none of the three is confined to it: 65 do-nothing episodes score above $0$, and $0.75$ is reached on 20 do-nothing and 23 spec-transcription tasks. No floor reaches $1.0$ on any task, while the reference arm resolves all 136.}
\label{fig:baselines}
\end{figure}

Table~\ref{tab:tasknames} maps each task name used in the text to its internal ID, which names the task's repository directory, and to its family. Table~\ref{tab:composition} gives the full per-family composition of the released corpus; Figure~\ref{fig:baselines} shows the score-lattice distributions of the four release-gate arms referenced in the main results.

\begin{table}[t]
\centering
\footnotesize
\setlength{\tabcolsep}{3pt}
\caption{Composition of FDE-Bench: 136 released tasks (66 greenfield, 70 repair). Mode: greenfield/repair task counts per family. Surfaces: single Docker images (37), Compose stacks (43), Kubernetes (56). The full-pipeline family (F, $n=5$) is reported as a case study, not as a quantitative axis.}
\label{tab:composition}
\begin{tabular}{@{}llp{3.1cm}cr@{}}
\toprule
Axis & Fam. & Description & G/R & Tasks \\
\midrule
Container. (C) & C1 & Author Dockerfiles from source, spec & 16/0 & 16 \\
 & C2 & Fix failing or non-conformant builds & 0/12 & 12 \\
 & C3 & Multi-container composition and wiring & 7/6 & 13 \\
 & C4 & Size, provenance, non-root, no secrets & 2/6 & 8 \\
\midrule
Orchestr. (O) & O1 & Manifest authoring from spec & 18/0 & 18 \\
 & O2 & Diagnose and fix live failures & 0/10 & 10 \\
 & O3 & Multi-service connectivity and discovery & 2/8 & 10 \\
 & O4 & Config/secret management to policy & 1/5 & 6 \\
 & O5 & Scale and rollout conformance & 2/2 & 4 \\
\midrule
Health (H) & H1 & Probes matching application health semantics & 6/7 & 13 \\
 & H2 & Required PromQL-checkable metric series & 4/5 & 9 \\
 & H3 & Alerts: fire under fault, else silent & 3/6 & 9 \\
 & H4 & Log pipeline wiring & 0/3 & 3 \\
\midrule
Full pipe. (F) & F & Containerize, orchestrate, monitor end-to-end & 5/0 & 5 \\
\midrule
Total & & & 66/70 & 136 \\
\bottomrule
\end{tabular}
\end{table}

\begin{table}[p]
\centering
\scriptsize
\setlength{\tabcolsep}{2pt}
\caption{Task names. Each of the 136 released tasks, sorted by the name used in the text, with its internal ID (the task's directory under \texttt{tasks/}) and its family from Table~\ref{tab:composition}. The mapping is also released as \texttt{tasks/NAMES.yaml}.}
\label{tab:tasknames}
\begin{tabular}{@{}llc@{\hspace{6pt}}llc@{}}
\toprule
Name & ID & Fam. & Name & ID & Fam. \\
\midrule
\task{add-catalog-api-healthcheck} & h1-11 & H1 & \task{fix-catalog-cache-reachability-k8s} & o3-01 & O3 \\
\task{alert-on-depot-restart-loops} & h3-08 & H3 & \task{fix-catalog-stuck-init-k8s} & o2-06 & O2 \\
\task{alert-on-quote-api-latency} & h3-09 & H3 & \task{fix-checkout-instance-down-alert} & h3-01 & H3 \\
\task{alert-on-voucher-api-failures} & h3-07 & H3 & \task{fix-checkout-session-store-k8s} & o2-05 & O2 \\
\task{autoscale-payments-api-k8s} & o5-04 & O5 & \task{fix-deckhand-live-editing} & c3-07 & C3 \\
\task{compose-customers-api-with-schema} & c3-11 & C3 & \task{fix-dispatch-api-health-status} & h1-02 & H1 \\
\task{compose-hit-counter-stack} & c3-01b & C3 & \task{fix-dispatch-api-slow-stop} & c2-09 & C2 \\
\task{compose-jobcatch-demo} & c3-10 & C3 & \task{fix-dispatch-board-empty-k8s} & o3-07 & O3 \\
\task{compose-orders-reporting-proxy} & c3-12 & C3 & \task{fix-docs-portal-log-labels} & h4-03 & H4 \\
\task{compose-orders-service-schema-step} & c3-08 & C3 & \task{fix-edge-gateway-routing-table-k8s} & o4-04 & O4 \\
\task{compose-service-manual-and-search} & c3-13 & C3 & \task{fix-edge-proxy-flapping-k8s} & o2-02 & O2 \\
\task{compose-stockroom-cold-start} & c3-02b & C3 & \task{fix-fleet-console-image-build} & c2-10 & C2 \\
\task{containerize-digest-svc} & c1-08 & C1 & \task{fix-fleet-console-unprivileged-user} & c4-04 & C4 \\
\task{containerize-dispatch-notifier} & c1-10 & C1 & \task{fix-gate-scan-scrape-targets} & h2-04 & H2 \\
\task{containerize-docs-site} & c1-03 & C1 & \task{fix-gateway-startup-permission-error} & c2-04 & C2 \\
\task{containerize-edge-gateway} & c1-11 & C1 & \task{fix-handover-api-shared-db} & c3-09 & C3 \\
\task{containerize-feedback-api-read-only} & c4-07 & C4 & \task{fix-incident-reminder-cadence} & h3-06 & H3 \\
\task{containerize-field-report-portal} & c1-05 & C1 & \task{fix-ingest-indexer-lookup-k8s} & o3-03 & O3 \\
\task{containerize-hookgate-amd64} & c1-16 & C1 & \task{fix-ingest-worker-readiness-k8s} & o2-08 & O2 \\
\task{containerize-inventory-api} & c1-01b & C1 & \task{fix-ingestd-plaintext-password} & c4-06 & C4 \\
\task{containerize-ledger-recon-health} & h1-10 & H1 & \task{fix-intake-front-door-exits} & c3-05 & C3 \\
\task{containerize-license-svc-no-token} & c4-08 & C4 & \task{fix-inventory-api-over-quota} & c4-03 & C4 \\
\task{containerize-metrics-shim} & c1-04 & C1 & \task{fix-label-service-unhealthy} & h1-09 & H1 \\
\task{containerize-orders-service} & c1-06 & C1 & \task{fix-lastmile-api-no-target-k8s} & h2-03 & H2 \\
\task{containerize-reporting-api-analytics} & c1-09 & C1 & \task{fix-ledger-api-new-namespace-k8s} & o4-02 & O4 \\
\task{containerize-settlement-batch-worker} & c1-07 & C1 & \task{fix-ledger-stack-forgets-data} & c3-06 & C3 \\
\task{containerize-settlements-api} & c1-13 & C1 & \task{fix-metering-api-recording-k8s} & o3-05 & O3 \\
\task{containerize-storefront} & c1-14 & C1 & \task{fix-newsroom-error-page-k8s} & o3-06 & O3 \\
\task{containerize-taskdeck-api-reuse-deps} & c2-13 & C1 & \task{fix-notes-api-ci-build} & c2-01b & C2 \\
\task{containerize-telemetry-relay} & c1-15 & C1 & \task{fix-notes-service-startup} & c2-03 & C2 \\
\task{containerize-usage-reporting-api} & c1-12 & C1 & \task{fix-ops-console-hardened-startup} & c4-05 & C4 \\
\task{deploy-analytics-db-scratch-disk-k8s} & o1-14 & O1 & \task{fix-orderflow-no-logs} & h4-01 & H4 \\
\task{deploy-billing-without-kills-k8s} & o1-04 & O1 & \task{fix-orders-api-503s-k8s} & h1-05 & H1 \\
\task{deploy-corridor-api-no-outage-k8s} & h1-14 & H1 & \task{fix-orders-api-metrics-scrape} & h2-01 & H2 \\
\task{deploy-coverage-web-default-deny-k8s} & o3-12 & O3 & \task{fix-orders-api-rollout-k8s} & o5-01 & O5 \\
\task{deploy-depot-inventory-stack} & f-01 & F & \task{fix-orders-never-ready-k8s} & o2-07 & O2 \\
\task{deploy-dispatch-health-checks-k8s} & h1-07 & H1 & \task{fix-paymentgw-stack-trace-rows} & h4-02 & H4 \\
\task{deploy-editable-status-page-k8s} & o1-03 & O1 & \task{fix-pick-list-blank-panels} & h2-06 & H2 \\
\task{deploy-engineering-index-k8s} & o1-09 & O1 & \task{fix-pricing-api-slim-startup} & c2-07 & C2 \\
\task{deploy-goods-intake-stack-k8s} & f-04 & F & \task{fix-pricing-oom-kills-k8s} & o2-09 & O2 \\
\task{deploy-handbook-nonroot-k8s} & o1-15 & O1 & \task{fix-pricing-site-bind-denied-k8s} & o2-10 & O2 \\
\task{deploy-https-storefront-with-bundle} & f-05 & F & \task{fix-queue-depth-alert} & h3-03 & H3 \\
\task{deploy-ingest-relay-readiness-k8s} & o1-13 & O1 & \task{fix-quote-api-restarts-k8s} & h1-08 & H1 \\
\task{deploy-ingest-worker-backlog-paging} & f-02 & F & \task{fix-reporting-api-not-permitted-k8s} & o4-03 & O4 \\
\task{deploy-intake-dependency-wait-k8s} & o1-16 & O1 & \task{fix-revenue-reporting-api-build} & c2-06 & C2 \\
\task{deploy-inventory-api-k8s} & o1-01 & O1 & \task{fix-riskscore-api-disclosed-token} & c4-02 & C4 \\
\task{deploy-metrics-agent-nonroot-k8s} & o1-11 & O1 & \task{fix-scan-backlog-paging} & h3-05 & H3 \\
\task{deploy-notifier-vendor-token-k8s} & o1-06 & O1 & \task{fix-search-api-office-network-k8s} & o3-02 & O3 \\
\task{deploy-ops-dashboard-identity-k8s} & o1-08 & O1 & \task{fix-shift-scheduler-restarts-k8s} & h1-06 & H1 \\
\task{deploy-orders-with-readiness-k8s} & o1-12 & O1 & \task{fix-shipping-quote-startup} & c2-08 & C2 \\
\task{deploy-orders-with-schema-once-k8s} & o1-07 & O1 & \task{fix-signing-gateway-migration-k8s} & o4-08 & O4 \\
\task{deploy-payments-api-credentials-k8s} & o4-06 & O4 & \task{fix-status-tools-crashloop-k8s} & o2-01 & O2 \\
\task{deploy-quote-batch-memory-k8s} & o1-17 & O1 & \task{fix-storefront-empty-shop} & c3-04 & C3 \\
\task{deploy-reports-never-silent-k8s} & o1-02 & O1 & \task{fix-storefront-endpoint-down-alert} & h3-02 & H3 \\
\task{deploy-staging-postgres-storage-k8s} & o1-05 & O1 & \task{fix-storefront-external-access-k8s} & o3-04 & O3 \\
\task{deploy-storefront-security-standard} & f-03 & F & \task{fix-storefront-reverse-proxy} & o3-08 & O3 \\
\task{deploy-sweep-scheduler-health-k8s} & h1-13 & H1 & \task{fix-storefront-web-ci-ship-k8s} & o2-03 & O2 \\
\task{deploy-tenant-portal-hardened-k8s} & o1-19 & O1 & \task{fix-tariff-engine-startup-flapping} & h1-03 & H1 \\
\task{deploy-tick-relay-health-config-k8s} & h1-12 & H1 & \task{fix-telemetry-relay-exits-on-start} & c2-11 & C2 \\
\task{deploy-transcoder-with-scratch-k8s} & o1-18 & O1 & \task{fix-webhook-relay-amd64-startup} & c2-12 & C2 \\
\task{export-jobproc-metrics} & h2-08 & H2 & \task{fix-webhook-signer-quarantine} & c4-01 & C4 \\
\task{fix-analytics-stack-clean-checkout} & c3-03 & C3 & \task{instrument-catalogue-api-metrics} & h2-07 & H2 \\
\task{fix-analytics-store-pending-k8s} & o2-04 & O2 & \task{instrument-ledger-api-busy-metrics} & h2-05 & H2 \\
\task{fix-auth-api-dies-on-signin} & c2-05 & C2 & \task{release-token-svc-without-drops-k8s} & o5-03 & O5 \\
\task{fix-billing-muted-alert-channel} & h3-04 & H3 & \task{rotate-fulfilment-db-credential-k8s} & o4-05 & O4 \\
\task{fix-billing-prometheus-says-down} & h2-02 & H2 & \task{scrape-every-dispatch-instance} & h2-09 & H2 \\
\task{fix-billing-worker-build} & c2-02b & C2 & \task{split-depot-compose-stacks} & o3-09 & O3 \\
\task{fix-broker-crashloop-k8s} & h1-04 & H1 & \task{unwedge-checkout-api-cutover-k8s} & o5-02 & O5 \\
\bottomrule
\end{tabular}
\end{table}

\section{Full-Pipeline Family as a Case Study}
\label{app:fullpipeline}

The five full-pipeline tasks combine all three axes in one episode and include checks across all four layers. At $n=5$, one task changes the resolved rate by 20 percentage points. The family is therefore excluded from the per-axis aggregates of the main results and reported as a case study here.

Four of the five tasks were resolved by at least one agent, but none by all seven. No agent resolved \task{deploy-goods-intake-stack-k8s}, a five-service Kubernetes pipeline with an idempotent-migration requirement; five of its seven episodes stalled at readiness. Agent agreement is lowest in this family: six agents resolve \task{deploy-ingest-worker-backlog-paging}, and three each resolve \task{deploy-storefront-security-standard} and \task{deploy-https-storefront-with-bundle}, with no containment among the solver sets.

The floor baselines also differ on these five tasks. The do-nothing arm scores 0.000 on all five, spec-transcription scores above zero on only one, and the generic stub never passes readiness. These results show limited partial credit for the tested floor strategies on full-pipeline tasks. Expanding this family is the top priority for the next task wave.

\section{Related-Benchmark Comparison}
\label{app:related}

Table~\ref{tab:related} compares FDE-Bench feature by feature with the closest benchmarks discussed in the main text.

\begin{table*}[t]
\centering
\footnotesize
\setlength{\tabcolsep}{3pt}
\caption{Closest benchmarks (\checkmark{} = first-class, \(\circ\) = partial, -- = absent) on grading against a machine-readable deployment specification, container artifacts, orchestration, monitoring, behavioral checks on the running system, repair tasks, and fully programmatic (judge-free) scoring.}
\label{tab:related}
\begin{tabular}{@{}lccccccc@{}}
\toprule
Benchmark & Spec & Containers & Orch. & Monitor & Behav. & Repair & No judge \\
\midrule
EnvBench \citep{eliseeva2025envbench} & -- & -- & -- & -- & -- & -- & \checkmark \\
SetupBench \citep{arora2025setupbench} & -- & -- & -- & -- & \(\circ\) & \(\circ\) & \checkmark \\
Repo2Run \citep{hu2025repo2run} & -- & \checkmark & -- & -- & \(\circ\) & -- & \checkmark \\
DeployBench \citep{wang2026deploybench} & \(\circ\) & -- & -- & -- & \(\circ\) & -- & \checkmark \\
AIOpsLab \citep{shetty2025aiopslab} & -- & -- & \(\circ\) & \(\circ\) & \checkmark & \checkmark & \checkmark \\
ITBench \citep{jha2026itbench} & -- & -- & \(\circ\) & \(\circ\) & \checkmark & \checkmark & \checkmark \\
DevOps-Gym \citep{tang2026devopsgym} & -- & -- & -- & \(\circ\) & \(\circ\) & \checkmark & \checkmark \\
IaC-Eval \citep{kon2024iaceval} & \(\circ\) & -- & -- & -- & -- & -- & \checkmark \\
Terminal-Bench \citep{tbench2026terminalbench} & -- & \(\circ\) & -- & -- & \(\circ\) & \(\circ\) & \checkmark \\
\midrule
\textbf{FDE-Bench (ours)} & \checkmark & \checkmark & \checkmark & \checkmark & \checkmark & \checkmark & \checkmark \\
\bottomrule
\end{tabular}
\end{table*}

DevOps-Gym \citep{tang2026devopsgym} evaluates build configuration, monitoring, issue resolution, and test generation for Java and Go projects, with both individual stages and combined workflows. Its tasks include configuration repair and implementation, and its monitoring stage asks an agent to diagnose an injected performance or resource anomaly from command-line tools rather than to configure monitoring. It excludes infrastructure management and CI/CD automation from scope, on the stated grounds that those stages depend on mutable external systems such as cloud APIs and Kubernetes clusters; containers are its execution environment rather than a graded artifact. FDE-Bench holds application code fixed and grades submitted Docker, Compose, and Kubernetes artifacts by rebuilding a deployment and checking its specification. CodeFuse's DevOps-Eval \citep{codefuse2023devopseval} is a project covering DevOps knowledge, AIOps, and tool-use evaluation.

\section{Failure Taxonomy}
\label{app:taxonomy}

Table~\ref{tab:taxonomy} defines 14 diagnostic categories for inspecting failed checks and execution logs. Quantitative results use the observed check-layer states and recorded termination statuses in Table~\ref{tab:stage_termination}.

\begin{table}[t]
\centering
\footnotesize
\setlength{\tabcolsep}{3pt}
\caption{Fourteen failure diagnostic categories, grouped by prefix: spec (S), build (B), orchestration (O), monitoring (M), agent behavior (A).}
\label{tab:taxonomy}
\begin{tabular}{@{}llll@{}}
\toprule
Code & Category & Code & Category \\
\midrule
S1 & Spec misread & O2 & Networking / discovery \\
S2 & Under-spec.\ mishandled & O3 & State wiring \\
B1 & Wrong base image / arch & M1 & Probe misconfiguration \\
B2 & Dependency resolution & M2 & Metrics / alert wiring \\
B3 & Dockerfile semantics & A1 & Gave up early \\
O1 & Manifest semantics & A2 & Thrashing loop \\
A3 & Check-gaming attempt & A4 & Budget-cap termination \\
\bottomrule
\end{tabular}
\end{table}

\paragraph{Episode termination.}
Table~\ref{tab:termination} accounts for all 952 valid graded episodes in the v1.1 grid. Step or wall-clock limits ended 226 episodes; 103 remained unresolved and 123 resolved on replay. Six episodes ended in output-token truncation (two GPT-5.6-Luna, one DeepSeek-V4-Pro, three DeepSeek-V4-Flash), all failing at build; one GPT-5.6-Terra episode has unknown termination status. These seven episodes retain their replay scores in the evaluation grid.

\begin{table}[t]
\centering
\footnotesize
\setlength{\tabcolsep}{4pt}
\caption{Termination statuses among valid graded episodes. Step and Time count \texttt{max\_steps} and \texttt{timeout}; Token counts output-token truncations. Unknown retains missing termination metadata. The five status columns sum to Valid; Step plus Time gives the total Caps count in the main table.}
\label{tab:termination}
\begin{tabular}{@{}lrrrrrr@{}}
\toprule
Model & Valid & Submitted & Step & Time & Token & Unknown \\
\midrule
Gemini-3.6-Flash & 136 & 94 & 42 & 0 & 0 & 0 \\
GPT-5.6-Terra & 136 & 133 & 2 & 0 & 0 & 1 \\
Claude-Sonnet-5 & 136 & 112 & 24 & 0 & 0 & 0 \\
DeepSeek-V4-Pro & 136 & 82 & 53 & 0 & 1 & 0 \\
GPT-5.6-Luna & 136 & 134 & 0 & 0 & 2 & 0 \\
DeepSeek-V4-Flash & 136 & 89 & 44 & 0 & 3 & 0 \\
Gemini-3.5-Flash-Lite & 136 & 75 & 60 & 1 & 0 & 0 \\
\midrule
Total & 952 & 719 & 225 & 1 & 6 & 1 \\
\bottomrule
\end{tabular}
\end{table}

\begin{table}[htbp]
\centering
\footnotesize
\caption{Failure stage by recorded termination status, covering all 313 unresolved valid episodes in the v1.1 grid. Cap combines step and wall-clock limits. Token denotes output-token truncation; Unknown retains missing metadata. Stages follow the reporting priority defined in the main text.}
\label{tab:stage_termination}
\begin{tabular}{@{}lrrrrr@{}}
\toprule
Failure stage & Submitted & Cap & Token & Unknown & Total \\
\midrule
Build & 24 & 22 & 6 & 0 & 52 \\
Readiness & 66 & 43 & 0 & 1 & 110 \\
Behavior & 64 & 33 & 0 & 0 & 97 \\
Conformance & 49 & 5 & 0 & 0 & 54 \\
\midrule
Total & 203 & 103 & 6 & 1 & 313 \\
\bottomrule
\end{tabular}
\end{table}

\section{Check-Layer Weights and Binary Scoring}
\label{app:weights}

This revision applies equal weights of 0.25 to the four gated binary layers (scoring version \texttt{fde-ds-equal-v2}). We recompute DS from retained layer verdicts for 952 model episodes and 653 floor/reference records. Historical raw scores use 0.20/0.25/0.35/0.20 and remain archived. The five floor/reference arms retain the same task sets; the template arm has 109 valid results. Reweighting preserves every resolution decision and, on the v1.1 grid, the overall model DS ordering. The original-weight means rank the seven models in the same order as the equal-weight means, from 0.851 for Gemini-3.6-Flash to 0.699 for Gemini-3.5-Flash-Lite, while some within-axis orderings and ties change.

All four layers are binary. An earlier pilot scored the behavior and conformance layers as per-check pass fractions. A single failed check then produced five different scores across seven tasks because authors had written different numbers of checks. This made cross-task scores depend on check counts. Under equal weights, binary layers give every task the same score lattice, $\{0, 0.25, 0.50, 0.75, 1.00\}$. The failure-stage breakdown reports the first failing layer in build, readiness, behavior, conformance order. This attribution priority keeps the categories disjoint while the last two scoring branches remain parallel. A failed prerequisite leaves downstream functional correctness unmeasured; partial DS records achieved deployment stages.

The gate products make build and readiness prerequisites for both behavior and conformance. Behavior and conformance are parallel requirements: conformance can earn credit even if behavior fails, and vice versa. Reweighting these states does not change the requirement that all four layers pass for resolution. Resolved Rate is therefore invariant to the layer weights, while the secondary Deployment Score depends on the credit assigned to intermediate states.

\section{Family-Clustered Sign-Flip Permutation Test}
\label{app:permtest}

Pairwise model comparisons in the main text use a family-clustered sign-flip permutation test on per-task resolved outcomes. For models $A$ and $B$ sharing $n$ graded tasks, let $d_i \in \{-1, 0, 1\}$ be the resolved-outcome difference on task $i$, and let the tasks partition into the 14 scenario families $f \in \mathcal{F}$. The observed statistic is the mean difference
\begin{equation}
\label{eq:permstat}
T \;=\; \frac{1}{n} \sum_{f \in \mathcal{F}} \sigma_f \sum_{i \in f} d_i, \qquad \sigma_f = 1,
\end{equation}
and the null distribution assigns each family's sign $\sigma_f$ independently and uniformly from $\{-1, +1\}$. Flipping whole families preserves the dependence among task variants that share workspace structure. The test assumes exchangeability of paired model labels at the family level.

The exact two-sided $p$-value is the fraction of the $2^{14}$ sign assignments whose $|T|$ meets or exceeds the observed value. The specified Monte Carlo procedure approximates this fraction using 20{,}000 sampled assignments with seed 900. An exact reanalysis enumerating all 16{,}384 assignments on the v1.1 grid gives $p = 0.0015$ (24 of the 16{,}384 assignments) for Gemini-3.6-Flash versus Gemini-3.5-Flash-Lite and $p = 0.59$ (9{,}728 of 16{,}384) versus Claude-Sonnet-5; the main results report these exact values.

\section{Claude-Opus-5 Probe Records}
\label{app:probe}

The probe used Anthropic's recorded model identifier \texttt{claude-opus-5}. All 16 attempts record harness version \texttt{1.0.0-rc1}, seed 900, and SDK version 0.93.0. Recorded start times span July 28--29, 2026 (UTC). The provider configuration specifies adaptive thinking with summarized display, high effort, a 16{,}000-token per-turn limit, and a 200{,}000-output-token episode limit. Temperature is stored as \texttt{null}; a dated server-side model snapshot is not recorded.

Table~\ref{tab:probe} lists the twelve recorded initial probes and all six additional task IDs. The corresponding pilot-selection ledger is unavailable. The initial set yielded eleven resolved episodes out of twelve, with no recorded cap hit. Of four additional attempts, three valid episodes resolved. The \task{deploy-https-storefront-with-bundle} attempt ended with \texttt{api\_error}; its artifacts passed replay, but the episode is excluded under the API-error rule. Thus the records contain 15 valid episodes with 14 resolutions, plus one API-interrupted attempt with a passing artifact. Two additional tasks have no probe record. These are historical selected-task results. Tasks \task{release-token-svc-without-drops-k8s} and \task{autoscale-payments-api-k8s} were excluded from the frozen 133-task grid and return in the 136-task v1.1 grid after a check-suite fix (Appendix~\ref{app:defectpolicy}); the probe episodes on them predate that fix, and no roster model resolves either task in the v1.1 grid.

\begin{table}[htbp]
\centering
\scriptsize
\setlength{\tabcolsep}{4pt}
\caption{Historical Claude-Opus-5 probe ledger. Valid and Resolved refer to episodes; DS grades the retained artifacts. A dash marks an absent probe or an inapplicable episode outcome. The last column indicates whether the task was in the frozen 133-task grid; all 18 tasks are in the 136-task v1.1 grid.}
\label{tab:probe}
\begin{tabular}{@{}lllcrrrc@{}}
\toprule
Set & Task & Termination & Valid & Resolved & DS & Steps & Frozen grid \\
\midrule
Initial & \task{containerize-digest-svc} & submitted & yes & yes & 1.0 & 23 & yes \\
 & \task{fix-ops-console-hardened-startup} & submitted & yes & yes & 1.0 & 23 & yes \\
 & \task{containerize-feedback-api-read-only} & submitted & yes & yes & 1.0 & 13 & yes \\
 & \task{containerize-license-svc-no-token} & submitted & yes & yes & 1.0 & 13 & yes \\
 & \task{deploy-storefront-security-standard} & submitted & yes & yes & 1.0 & 46 & yes \\
 & \task{instrument-ledger-api-busy-metrics} & submitted & yes & no & 0.0 & 32 & yes \\
 & \task{fix-scan-backlog-paging} & submitted & yes & yes & 1.0 & 18 & yes \\
 & \task{fix-incident-reminder-cadence} & submitted & yes & yes & 1.0 & 19 & yes \\
 & \task{alert-on-depot-restart-loops} & submitted & yes & yes & 1.0 & 30 & yes \\
 & \task{deploy-ops-dashboard-identity-k8s} & submitted & yes & yes & 1.0 & 5 & yes \\
 & \task{deploy-coverage-web-default-deny-k8s} & submitted & yes & yes & 1.0 & 17 & yes \\
 & \task{release-token-svc-without-drops-k8s} & submitted & yes & yes & 1.0 & 5 & no \\
\midrule
Additional & \task{compose-service-manual-and-search} & submitted & yes & yes & 1.0 & 23 & yes \\
 & \task{deploy-goods-intake-stack-k8s} & -- & -- & -- & -- & -- & yes \\
 & \task{deploy-https-storefront-with-bundle} & \texttt{api\_error} & no & -- & 1.0 & 29 & yes \\
 & \task{split-depot-compose-stacks} & submitted & yes & yes & 1.0 & 29 & yes \\
 & \task{fix-signing-gateway-migration-k8s} & -- & -- & -- & -- & -- & yes \\
 & \task{autoscale-payments-api-k8s} & submitted & yes & yes & 1.0 & 13 & no \\
\bottomrule
\end{tabular}
\end{table}

\section{Disclosed Task Defects and Grid Membership}
\label{app:defectpolicy}

A task leaves the grid under one rule: its current instance cannot yield a valid verdict for any model. Two conditions meet it. The reference solution stops resolving the task under re-verification, because a task the reference cannot solve has no demonstrated solution path. The check suite itself fails to execute. Both conditions void every model's verdict on the task at once, so removal loses no comparative information. The frozen 133-task grid excluded three tasks under this rule; the 136-task v1.1 grid excludes none, because all three returned after their defects were repaired. Task \task{deploy-engineering-index-k8s} returned after its packaged reference manifest regained the foreground flag (\texttt{-f}) of its \texttt{httpd} command; the 2026-07-28 packaging had dropped the flag, and without it the reference had stopped resolving. Tasks \task{release-token-svc-without-drops-k8s} and \task{autoscale-payments-api-k8s} returned after a check-suite thread-cleanup fix. In the frozen suite an event named \texttt{\_stop} had shadowed the thread method and made \texttt{join()} raise \texttt{TypeError}. All three were re-gated and re-run for every roster model; six of the seven models resolve \task{deploy-engineering-index-k8s}, and none resolves \task{release-token-svc-without-drops-k8s} or \task{autoscale-payments-api-k8s}.

\paragraph{The v1.1 re-run.} The v1.1 grid keeps the 136 released tasks. Twenty-nine tasks (twelve in C2, \task{fix-handover-api-shared-db}, seven in C4, \task{fix-tariff-engine-startup-flapping}, \task{containerize-ledger-recon-health}, and seven in H3) gained a stated stability-window clause in \texttt{spec.yaml} because their checks already graded it, and the shared Kubernetes workload-discovery rule was corrected (discovery version 4, anchored on the port or in-cluster URL the contract states, with several owners allowed). Every roster model was run once more on the 29 changed tasks and the three returning tasks, 224 episodes with seed 900 under the frozen harness, prompt, tools, and caps; together with the re-rolled \task{export-jobproc-metrics} episode described below, these 225 fresh episodes replace the corresponding frozen cells, so the grid holds 727 frozen and 225 v1.1 cells. The OpenAI-compatible adapter prices requests only from a supplied price table. The four Gemini and GPT models used per-token tables reconstructed from the frozen cost records: Gemini-3.6-Flash 1.50/0.15/7.50, Gemini-3.5-Flash-Lite 0.30/0.03/2.50, GPT-5.6-Terra 2.50/0.25/15.00, and GPT-5.6-Luna 1.00/0.10/6.00 USD per million uncached-input, cached-input, and output tokens. The 20 episodes generated before the tables were passed carry a recorded cost of zero and are repriced from their token counts in a sidecar file. The re-run cost 28.1 USD with that repricing (25.5 USD recorded). Two invalid episodes were re-rolled once each: the Gemini-3.6-Flash \task{export-jobproc-metrics} episode, invalid in the frozen grid after a host-port collision, and the fresh Gemini-3.6-Flash \task{fix-ingestd-plaintext-password} episode, which ended in an API timeout. The release gate was re-run on the 29 changed tasks and on 35 Kubernetes tasks under discovery version 4 (\texttt{runs/\_regate\_v11.jsonl} and \texttt{runs/\_regate\_a\_family.jsonl}); on every task the reference and second solutions resolve, and the do-nothing, spec-transcription, generic-stub, and three adversarial arms do not. The v1.0 records of every re-gated arm and every replaced model cell remain archived beside the v1.1 records.

A ticket that under-describes the grading procedure does not meet that rule, because it leaves the verdicts intact and affects models unequally. Task \task{instrument-ledger-api-busy-metrics} is the one such case in the grid. Its verification text describes build and deploy but omits that replay pulls the submitted image from its local tag, and this stage is where two episodes ended. Table~\ref{tab:h205} gives the per-model outcome. Six of the seven roster models cleared the build layer and three resolved the task, so the omission did not make the task unsolvable; it coincides with failure only for the two Anthropic models, whose episodes scored zero at build. The task therefore stays in the grid, its two build-stage failures are recorded in the failure ledger as attributable to the disclosure defect, and the effect of removing it is reported here rather than applied.

\begin{table}[t]
\centering
\footnotesize
\caption{Per-model outcome on \task{instrument-ledger-api-busy-metrics} and the effect of removing it. Layers passed are the gated binary layers the episode cleared; the disclosed defect sits in the build layer. The diagnostic rate removes \task{instrument-ledger-api-busy-metrics} for every model, leaving all other v1.1 grid episodes unchanged; it is a sensitivity check, not the reported analysis. Claude-Opus-5 appears as a probe record and is not on the roster.}
\label{tab:h205}
\begin{tabular}{@{}llcrrr@{}}
\toprule
Model & Task outcome & Layers passed & DS & RR \% & Diagnostic RR \% \\
\midrule
Gemini-3.6-Flash & resolved & all four & 1.00 & 75.00 & 74.81 \\
GPT-5.6-Terra & resolved & all four & 1.00 & 69.85 & 69.63 \\
GPT-5.6-Luna & resolved & all four & 1.00 & 63.24 & 62.96 \\
DeepSeek-V4-Pro & behavior failed & build, ready, conformance & 0.75 & 69.85 & 70.37 \\
DeepSeek-V4-Flash & behavior failed & build, ready, conformance & 0.75 & 66.18 & 66.67 \\
Gemini-3.5-Flash-Lite & behavior failed & build, ready, conformance & 0.75 & 52.94 & 53.33 \\
Claude-Sonnet-5 & build failed & none & 0.00 & 72.79 & 73.33 \\
\midrule
Claude-Opus-5 (probe) & build failed & none & 0.00 & -- & -- \\
\bottomrule
\end{tabular}
\end{table}

Removing \task{instrument-ledger-api-busy-metrics} uniformly moves every rate by less than one percentage point and leaves the best model unchanged. It does separate the one tie in the grid: DeepSeek-V4-Pro and GPT-5.6-Terra, both at 95 of 136, move to 70.37 against 69.63. The separation is not supported by a test, and the reported analysis retains the task.

\paragraph{Declared specification gaps.} The frozen release rubric marked 63 graded behaviors that no specification clause states, across 57 tasks: a single-workload precondition used by the check that discovers what to probe (28), a stability window requiring the deployment to stay up and report healthy for a fixed interval (29), and six isolated cases. Version 1.1 closes 56 of them. The 29 stability windows are now stated in the \texttt{spec.yaml} of the tasks whose checks grade them, and discovery version 4 anchors workload discovery on the port or in-cluster URL the contract states, which closes 27 of the 28 single-workload declarations (ledger: \texttt{paper/reviews/cr01-gap-closure-2026-09-23.md}). Seven declared gaps remain, on four tasks: the six isolated cases, namely two each on \task{fix-quote-api-restarts-k8s} and \task{deploy-corridor-api-no-outage-k8s} (a 404 status and a PID-1 command line that the specification does not name), one on \task{fix-storefront-endpoint-down-alert} (scrape-target health), and one on \task{deploy-intake-dependency-wait-k8s}, together with the one single-workload declaration that discovery version 4 leaves open, also on \task{deploy-intake-dependency-wait-k8s}; both \task{deploy-intake-dependency-wait-k8s} checks use task-local discovery that version 4 does not reach. Scanning every valid roster episode in the v1.1 grid for failures of these seven checks finds two episodes, DeepSeek-V4-Pro and DeepSeek-V4-Flash on \task{deploy-corridor-api-no-outage-k8s}, both frozen cells. Each also fails 11 checks that a specification clause does state, and no gated layer fails on gap checks alone, so dropping the seven checks would change no resolution and no Deployment Score. Two limits apply. Episodes are matched to the current rubric by task name, and a frozen episode may have run against a different task version. A gap that never fails an episode is still a contract defect, since a task whose grading asks for something its ticket does not state can mislead an agent without leaving a failing check behind.

\section{Resolved Rate by Axis and Mode}
\label{app:group_rr}

Table~\ref{tab:group_rr} includes the five full-pipeline tasks in the mode columns only. Every model has 136 valid episodes, so the axis denominators are 49, 48, and 34 and the mode denominators 66 and 70. The macro-average repair--greenfield gap is 30.7 percentage points.

\begin{table}[htbp]
\centering
\footnotesize
\setlength{\tabcolsep}{3pt}
\caption{Resolved Rate (\%) by axis and mode. Model entries give RR and resolved/valid episode counts. The macro mean weights the seven models equally.}
\label{tab:group_rr}
\begin{tabular}{@{}lccccc@{}}
\toprule
 & \multicolumn{3}{c}{RR by axis} & \multicolumn{2}{c}{RR by mode} \\
\cmidrule(lr){2-4} \cmidrule(lr){5-6}
Model & C & O & H & Greenfield & Repair \\
\midrule
Gemini-3.6-Flash & 73.5 (36/49) & 72.9 (35/48) & 82.4 (28/34) & 62.1 (41/66) & 87.1 (61/70) \\
GPT-5.6-Terra & 69.4 (34/49) & 64.6 (31/48) & 79.4 (27/34) & 59.1 (39/66) & 80.0 (56/70) \\
Claude-Sonnet-5 & 77.6 (38/49) & 68.8 (33/48) & 79.4 (27/34) & 57.6 (38/66) & 87.1 (61/70) \\
DeepSeek-V4-Pro & 67.3 (33/49) & 68.8 (33/48) & 76.5 (26/34) & 50.0 (33/66) & 88.6 (62/70) \\
GPT-5.6-Luna & 61.2 (30/49) & 60.4 (29/48) & 73.5 (25/34) & 51.5 (34/66) & 74.3 (52/70) \\
DeepSeek-V4-Flash & 71.4 (35/49) & 58.3 (28/48) & 70.6 (24/34) & 42.4 (28/66) & 88.6 (62/70) \\
Gemini-3.5-Flash-Lite & 49.0 (24/49) & 62.5 (30/48) & 50.0 (17/34) & 36.4 (24/66) & 68.6 (48/70) \\
\midrule
Macro mean & 67.1 & 65.2 & 73.1 & 51.3 & 82.0 \\
\bottomrule
\end{tabular}
\end{table}

\section{Historical Run Configuration}
\label{app:run_config}

\paragraph{Model requests.}
Table~\ref{tab:sdk} accounts for provider metadata in all 952 valid roster episodes of the v1.1 grid; the 225 v1.1 episodes record OpenAI SDK 2.30.0 or Anthropic SDK 0.93.0, and the frozen episodes record OpenAI SDK 2.30.0 or 2.50.0 or Anthropic SDK 0.120.2. Recorded configurations specify 16{,}000 output tokens per turn and 200{,}000 per episode. Temperature is stored as \texttt{null}, meaning that the client omitted it and used the provider default. Requests also omitted top-p, frequency and presence penalties, and API sampling seed. Seed 900 identifies the harness episode and seeded task probes; it does not fix provider sampling. Claude-Sonnet-5 used adaptive thinking with summarized display and high effort. The OpenAI-compatible client included a fallback that sent \texttt{reasoning\_effort=none} after a rejection. Effective server defaults and dated model snapshots were not retained.

\begin{table}[htbp]
\centering
\footnotesize
\caption{SDK versions in retained provider configurations (episode counts, v1.1 grid). The OpenAI-compatible client served OpenAI, DeepSeek, and Gemini models. One episode lacks provider configuration and retains its score.}
\label{tab:sdk}
\begin{tabular}{@{}lrrrrr@{}}
\toprule
Model & OpenAI 2.30.0 & OpenAI 2.50.0 & Anthropic 0.93.0 & Anthropic 0.120.2 & Missing \\
\midrule
Gemini-3.6-Flash & 41 & 95 & 0 & 0 & 0 \\
GPT-5.6-Terra & 43 & 92 & 0 & 0 & 1 \\
Claude-Sonnet-5 & 0 & 0 & 32 & 104 & 0 \\
DeepSeek-V4-Pro & 82 & 54 & 0 & 0 & 0 \\
GPT-5.6-Luna & 47 & 89 & 0 & 0 & 0 \\
DeepSeek-V4-Flash & 73 & 63 & 0 & 0 & 0 \\
Gemini-3.5-Flash-Lite & 39 & 97 & 0 & 0 & 0 \\
\bottomrule
\end{tabular}
\end{table}

\paragraph{Host and scheduling.}
The server handoff records 192 cores and 503 GB RAM; the adaptation log records 384 threads. Other workloads shared this host. The package also retains earlier episodes from a 16-thread laptop. CPU model, storage, full OS/runtime snapshots, and a complete per-episode host map were not archived. The server grading driver shards task IDs over 12 single-node kind clusters and limits each cluster to two simultaneous grading episodes. Separate per-episode namespaces (\texttt{fde-<run\_id>}) isolate object names; the grader rejects cluster-scoped resources and foreign namespaces and deletes each namespace after grading. Host probes use OS-allocated ephemeral ports through \texttt{kubectl port-forward}.

\paragraph{Network and budgets.}
The agent proxy allowlists container images, OS repositories, and language package repositories, including Debian/Ubuntu, Alpine, PyPI, npm, Go, Maven, and Rust. The historical audit observed PyPI access and blocked GitHub access. Grader-side image builds used unrestricted egress. Tool-call limits are fixed task-contract values shared across models: the 136-task v1.1 grid has 114 tasks at 40 calls, 14 at 45, four at 50, three at 55, and one at 35. The retained records contain no formula for deriving these values from task complexity.

\paragraph{Source identity.}
Reproduction requires harness commit \texttt{e2f4f48604c413791db4\allowbreak defbda58760a53e9093c} together with the archived inherited and run-specific patches. Table~\ref{tab:sdk} reports retained episode metadata.

\section{Reference Replay Coverage}
\label{app:replay_coverage}

Across 90 archived reference replays (30 development tasks; seeds 900, 901, and 902), each task has identical named check outcomes. For 23 tasks, seed 900's task fingerprint differs from seeds 901/902. The remaining seven (\task{deploy-editable-status-page-k8s}, \task{deploy-orders-with-schema-once-k8s}, \task{fix-storefront-web-ci-ship-k8s}, \task{fix-ingest-indexer-lookup-k8s}, \task{fix-dispatch-board-empty-k8s}, \task{fix-edge-gateway-routing-table-k8s}, and \task{unwedge-checkout-api-cutover-k8s}) match across all three runs. Thus 30 matching-fingerprint pairs plus seven third runs provide 67 runs and 1{,}138 check observations (392 checks twice and 118 three times). Fingerprints cover the task contract, specification, and task-local checker; full submitted-artifact and shared-grader identities were not retained. Reference checks that failed under host contention passed on an idle host.

We also replayed the current \task{containerize-metrics-shim} reference submission in three fresh deployments under \texttt{fde-ds-equal-v2}, labeled 900, 901, and 902. Full task-tree, shared-grader, and selected-artifact SHA-256 hashes match before and after all three runs. Each deployment passes all 13 named checks: 39 matching observations, with DS~1.0 per run.

\end{document}